\documentclass [11pt, letterpaper]{article}
\pdfoutput=1 
\usepackage{fullpage}
\usepackage{amsmath}
\usepackage{amssymb}
\usepackage{graphicx}
\usepackage{mathrsfs}
\usepackage{wrapfig}
\usepackage{boxedminipage}
\usepackage{epsfig}
\usepackage{setspace}
\usepackage{subfigure}
\usepackage{float}
\usepackage{hyperref}
\usepackage{enumerate}
\usepackage{cite}
\usepackage[font=footnotesize,labelfont=rm]{caption}

\begin{document}

\begin{flushright}
{\tt arXiv:1406.4533}
\end{flushright}

{\flushleft\vskip-1.35cm\vbox{\includegraphics[width=1.25in]{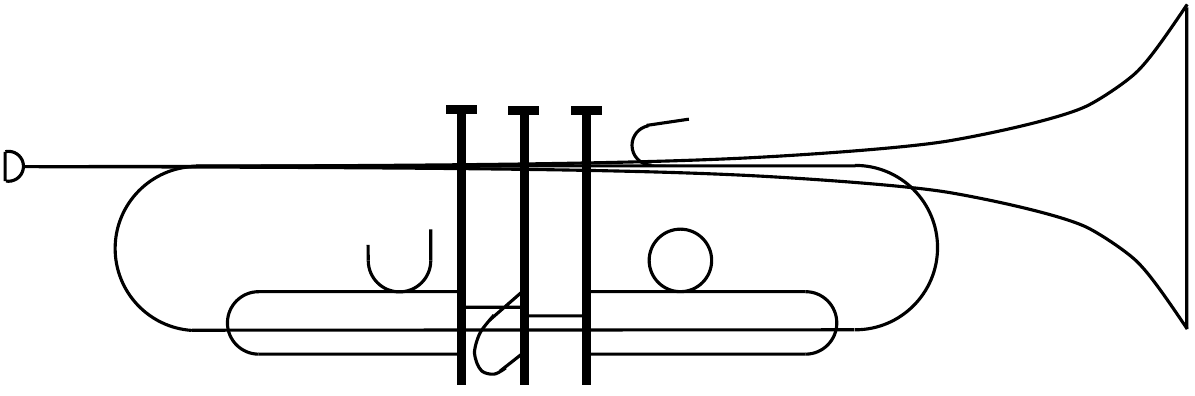}}}

\bigskip
\bigskip
\bigskip
\bigskip

\bigskip
\bigskip
\bigskip 
\begin{center} 

{\Large\bf The Extended Thermodynamic Phase Structure  }

\bigskip

{\large\bf of  }
\bigskip

{\Large\bf Taub--NUT and Taub--Bolt}

\end{center}

\bigskip \bigskip \bigskip \bigskip

\centerline{\bf Clifford V. Johnson}

\bigskip
\bigskip
\bigskip

  \centerline{\it Department of Physics and Astronomy }
\centerline{\it University of
Southern California}
\centerline{\it Los Angeles, CA 90089-0484, U.S.A.}

\bigskip

\centerline{\small \tt johnson1,  [at] usc.edu}

\bigskip
\bigskip


\begin{abstract} 
\noindent  We study aspects of the  extended gravitational thermodynamics  for the Taub--NUT and Taub--Bolt geometries in four dimensional locally anti--de Sitter spacetime, where the cosmological constant is treated as a dynamical pressure.  Attention is paid  to the phase structure in the $(p,T)$ plane, which has a line of first order phase transitions extending from the origin.  We argue for a  dynamical interpretation of the unstable physics in the negative specific heat region. A  deformation corresponding to a dyonic Taub--NUT/Bolt  system is also explored, and the effect of the deformation on the  phase diagram is characterised.

\end{abstract}
\newpage \baselineskip=18pt \setcounter{footnote}{0}

\section{Motivation}
%

Semi--classical quantum gravity has its most beautiful realisation in the context of black hole physics, where it breathes life into the equations describing their thermodynamics\cite{Bekenstein:1973ur,Bekenstein:1974ax,Hawking:1974sw,Hawking:1976de}. The standard central quantities are the black hole's mass $M$, surface gravity $\kappa$, and area $A$, which are related to the thermodynamic energy $U$, temperature $T$, and entropy $S$, respectively.
An odd (but perfectly consistent) feature of this framework is the absence of a variable representing pressure and its conjugate, volume, two central players in traditional discussions of thermodynamics when applied to everyday substances.  Recent work\footnote{For a selection of references, see refs.\cite{Caldarelli:1999xj,Wang:2006eb,Sekiwa:2006qj,LarranagaRubio:2007ut,Kastor:2009wy,Dolan:2010ha,Cvetic:2010jb,Dolan:2011jm,Dolan:2011xt}, including the reviews in refs.\cite{Dolan:2012jh,Altamirano:2014tva}. See also the early work in refs.\cite{Henneaux:1984ji,Teitelboim:1985dp,Henneaux:1989zc}.} 
has shown that the effective cosmological constant,~$\Lambda$, of the spacetime in question can  be treated as a pressure {\it via} $p=-\Lambda/8\pi G$, ($G$ is Newton's constant, and we will choose units such that  $c=1$, $\hbar=1$ and $k_{\rm B}=1$), and study of the resulting ``extended thermodynamics'' results in the natural suggestion\cite{Kastor:2009wy} that  black hole mass $M$  determines not  the internal energy $U$ but  the {\it enthalpy}: $M/G=H\equiv U+pV$, which includes a contribution from the energy of formation of the system. The thermodynamic volume $V$  can then be deduced in terms of the variables of the black hole spacetime in question. 

One interesting consequence of this is the observation that $V$ is not necessarily a geometrical volume defined in the spacetime at all, a fact revealed by {\it e.g.,} studying rotating black holes\cite{Cvetic:2010jb}. Motivated by the AdS/CFT correspondence\cite{Maldacena:1997re,Gubser:1998bc,Witten:1998qj,Witten:1998zw}, and by the fact that consistent traditional thermodynamics can be defined for non--black--hole spacetimes (see {\it e.g.} refs.\cite{Hawking:1998jf,Hawking:1976jb}), the suggestion was made in  ref.\cite{Johnson:2014xza}  to enlarge the framework of extended thermodynamics  to include {\it any} spacetime, again equating the mass to the enthalpy. The result was a larger class of examples of thermodynamic volumes which are non--geometrical (in the above sense), now  associated to the Taub--NUT and Taub--Bolt spacetimes\footnote{See also ref.\cite{MacDonald:2014zaa} for an example mixing the features of refs.\cite{Cvetic:2010jb,Johnson:2014xza} by combining nut charge and rotation.}. The  extreme example of Taub--NUT is especially notable for having a thermodynamic volume while there is no geometrical candidate at all.  We summarise some of the results we will need in sections~2 and~3.

In a sense, the extended thermodynamics brings the physics of black holes  closer to the physics of everyday (non--gravitational) substances  in a manner complementary to way  the AdS/CFT correspondence  does\cite{Witten:1998qj,Witten:1998zw}. For example, using the standard treatment of the thermodynamics, it was discovered\cite{Chamblin:1999tk,Chamblin:1999hg} that the fixed charge ensemble of Reissner--Nordstrom black holes in AdS has a rich thermodynamic phase structure reminiscent of the van der Waals description of a liquid/gas system, including a line of first order phase transitions ending in a second order critical point. Through AdS/CFT this maps holographically to a strongly coupled fluid in an unusual universality class. 
This is of course an exciting and important way of  defining non--traditional universality classes of fluids and other substances,  given its potential   use for interpreting and modelling new experimental discoveries. This is an idea that is well over a decade old\cite{Johnson:2003gi,Johnson:2010zzb}.  On the other hand it is intriguing to learn that black hole physics can be mapped (at least in part) to more traditional classes of substance, which is precisely what the extended thermodynamics has made possible.  For example, that same charged black hole system,  once pressure and volume are allowed to be dynamical, turns out to have the same critical properties as the van der Waals system\footnote{See ref.\cite{Kubiznak:2012wp} and also refs.\cite{Dolan:2011xt,Caldarelli:1999xj}.} including the standard mean field exponents. Several other familiar  thermodynamic phase diagram phenomena  (many in traditional universality classes) have now been mapped to black hole physics in the extended thermodynamics. Work on this continues by several groups\footnote{For recent reviews, see refs.\cite{Kubiznak:2014zwa,Altamirano:2014tva}.}. 

The meaning of the extended thermodynamics and what it implies for systems described holographically by AdS/CFT  remains to be spelled out. It is likely that the two areas can enrich each other once the right interpretation is found. As suggested in ref.\cite{Johnson:2014yja}, it seems clear that the natural AdS/CFT context where the extended thermodynamics might play a role would concern situations where there is  flow between field theories, perhaps involving operators that can dynamically change the effective degrees of freedom. Perhaps the extended thermodynamics is a good language for discussing dynamics on the space of field theories itself, and in the AdS/CFT language this boils down to dynamics on the space of holographic RG flows, a quite natural object in gauged supergravity, and perhaps string theory. Intriguingly, natural thermodynamic principles, such as the Second Law of thermodynamics (realised in terms of thermodynamic cycles representing heat engines and refrigerators) would translate into statements about the kinds of processes one could do in the space of field theories\footnote{This suggestion of ref.\cite{Johnson:2014yja} was based on the observation that the cosmological constant (and hence $p$) is connected to the number of degrees of freedom in the dual field theory by virtue of being related to the rank, $N$, of the dual gauge theory. Note that although they did not pursue the connection, refs.\cite{Dolan:2013dga,Kastor:2009wy} also mention the link between $p$ and $N$, and wondered as to its significance. We thank B. Dolan and D. Kastor for pointing this out.}. 

One of the other points made in ref.\cite{Johnson:2014yja} is that the enlarged dynamical setting of the extended thermodynamics helps  place sometimes puzzling thermodynamic properties of spacetimes into a light where they can be seen to make physical sense. For example, AdS$_4$ with an $S^3$ slicing (where Euclidean time is an $S^1$ fibred over an $S^2$) has negative entropy. For fixed cosmological constant this is simply strange. On the other hand, AdS$_4$ is a perfectly good spacetime that has just been sliced unusually. Rather than discard it, it would seem prudent to simply find a context in which this slicing makes sense. Putting it into the extended thermodynamics context and using the enthalpy prescription reveals that it has a negative thermodynamic volume\cite{Johnson:2014yja}, which might also seem odd. However, in a context where there are dynamically changing pressures and volumes, and accompanying heat flow,   both work done and heat flow into the volume naturally come with both signs, and so for positive $T$ and $p$ can result in negative $S$ and $V$, with the correct interpretation.

 A negative specific heat should have a natural dynamical narrative as well.  It is familiar, for example in the case of Schwarzschild black holes in flat space. In extended thermodynamics it becomes the specific heat at constant pressure\footnote{The counterpart specific heat, $C_V$, vanishes for static black holes\cite{Dolan:2010ha}.}, $C_p$. Negative specific heat suggests a thermal instability, and so we are strictly outside equilibrium thermodynamics now, but can at least  start to follow the physics.  The system runs to higher  temperature (like a flat space black hole radiating away its energy), lowering its entropy. 
 This is a rather runaway process for a black hole in flat space, and so a discussion of it is somewhat incomplete since it runs off to regimes that lie well outside the semi--classical context in which we began. Adding a cosmological constant avoids the problem altogether, rather than solving it: Analogues of the Schwarzschild black holes exist in AdS as the solutions representing the ``small'' branch of black holes, but they never appear in the phase diagram, (either in the standard or extend thermodynamics). At any pressure (and hence $\Lambda$) the low temperature phase is AdS (with the traditional $S^1\times S^2$ slicing), and then there is the Hawking--Page transition\cite{Hawking:1982dh} temperature above which the ``large'' branch black holes are thermodynamically favoured. So the negative $C_p$ of the black holes never needs to be interpreted. 
 
 Nevertheless, it seems natural to try to give negative $C_p$ an interpretation in other systems where it might not be avoided, with the help of the extended thermodynamics.  We have a context in which to do so: Start again with AdS$_4$ with the $S^3$ slicing. It lies on an extended phase diagram that we will explore more in this paper. We are free to adjust the temperature of this slicing by viewing AdS$_4$ as part of the Taub--NUT--AdS class of spacetimes\cite{Chamblin:1998pz}, characterised by nut charge $n$, with temperature $T=1/8\pi n$. The case $n=\ell/2$ (where $\ell$ is the length scale set by the cosmological constant $\Lambda=-3/\ell^2$) is AdS$_4$. The Taub--NUT--AdS spaces have\cite{Mann:1999pc,Emparan:1999pm} a characteristic temperature, $T^o$, above which $C_p<0$.\footnote{They also have lower temperature below which  $S<0$, but  we will not concern ourselves much with this region of parameter space.} 
At another temperature $T^*$ above $T^o$,  there is a transition  to the (``large'' branch) Taub--Bolt spacetime, which is stable\cite{Chamblin:1998pz}.  Notice that this leaves a window $T^o<T<T^*$  where the geometry with lower action has $C_p<0$. This demands to be treated as interpretable sensible physics, and we do so. It is in a sense a ``safer'' system than the flat space Schwarzschild black hole case since the physics  does not run away to regimes outside the semi--classical domain; the Taub--bolt spacetime takes over, restoring  the equilibrium thermodynamics.

It is in this spirit we present and discuss the $(p,T)$ phase diagram of the Taub--NUT/Taub--Bolt  system, in section~4. It is also natural to consider a deformation of our system by electric and magnetic fields that make the NUT/Bolt solutions into dyons. The electric and magnetic charges of the solutions (discussed in section~5) will depend upon one parameter, $v$, and it will be natural to study both real and imaginary values of this parameter, unpacking the thermodynamic phase structure. We discuss the resulting  phase diagrams in section~6.

\section{Taub--NUT--AdS and  Taub--Bolt--AdS}
We work with Einstein--Hilbert action:
\begin{equation}
I=-\frac{1}{16\pi G}\int \! d^4x \sqrt{-g} \left(R-2\Lambda\right)\ ,
\label{eq:action}
\end{equation}
with negative  cosmological constant $\Lambda$ which sets a length scale $\ell$:  $\Lambda=-3/\ell^2$. 
The Taub--NUT \cite{Taub:1950ez,Newman:1963yy} and Taub--Bolt \cite{Page:1979aj} spacetimes,  have a metric of the following form for negative cosmological constant\cite{Page:1985bq,Page:1985hg}:
\begin{equation}
\label{eq:metric}
ds^2=F(r)(d\tau+2n\cos\theta d\phi)^2+\frac{dr^2}{F(r)}+(r^2-n^2)(d\theta^2+\sin^2\theta d\phi^2)\ ,
\end{equation}
where
\begin{equation}
\label{eq:metric-function}
F(r)\equiv\frac{(r^2+n^2)-2mr+\ell^{-2}(r^4-6n^2r^2-3n^4)}{r^2-n^2}\ .
\end{equation}
Time $\tau$ here is Euclidean with period $\beta=8\pi n$ (in order to ensure the physical absence of Misner strings\cite{Misner:1963fr}). The spaces are locally asymptotically AdS$_4$, generically; The topological $S^3$ formed by the $\tau$ circle fibred over the $S^2$ of $\theta$ and $\phi$ is squashed for general $n$.  The case $n=\ell/2$  in fact yields AdS$_4$, with radial slices that are round  $S^3$s\cite{Chamblin:1998pz}, as discussed in the previous section. 

\begin{wrapfigure}{R}{0.45\textwidth}
{\centering
\includegraphics[width=2.5in]{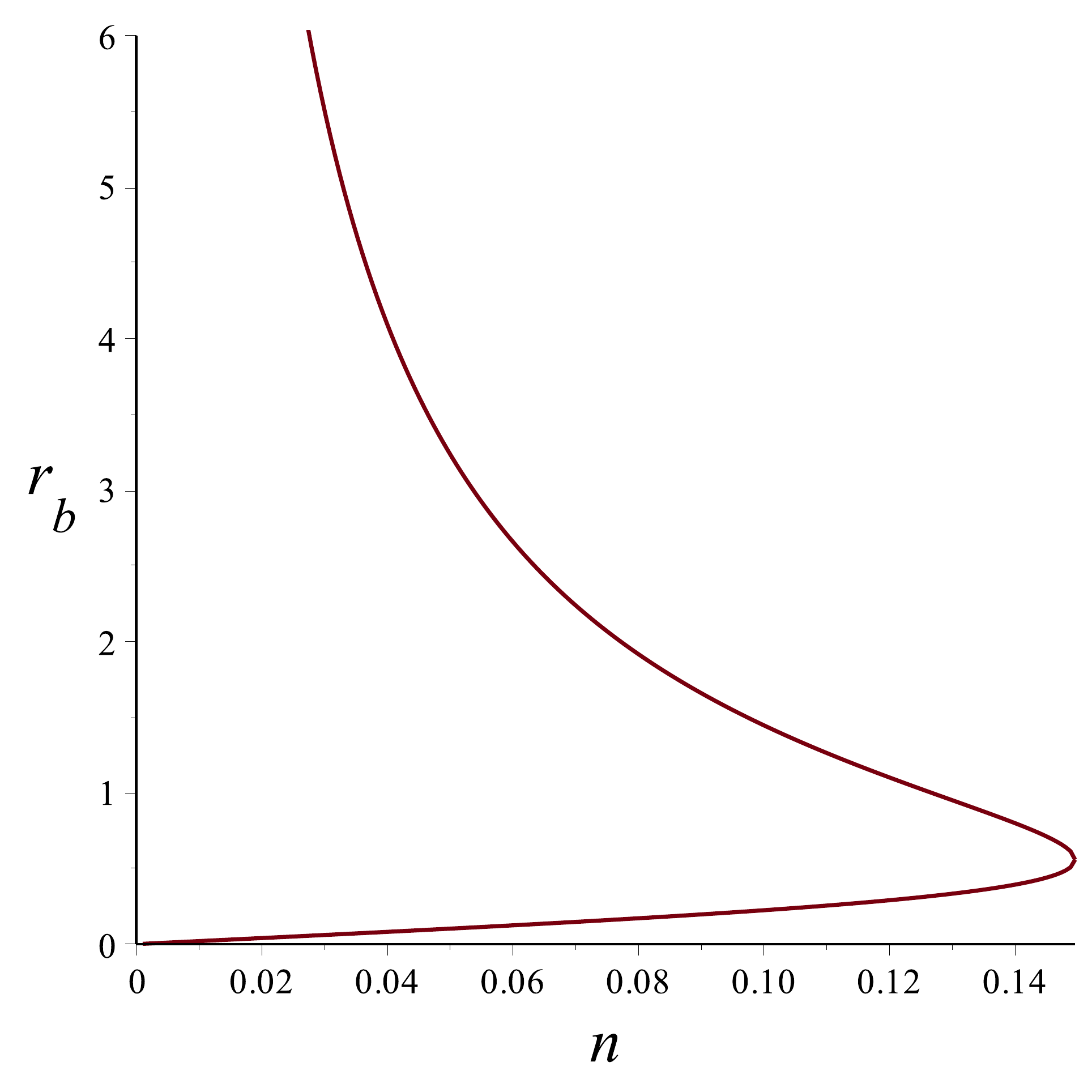} 
   \caption{\footnotesize   The available bolt radii, the union of an upper and a lower branch that connect at $n_{\rm max}\simeq 0.149\ell$. Here we chose $\ell=1$.}  \label{fig:arbyplot}
}
\end{wrapfigure}
The circle parameterized by  $\tau$ is Hopf fibred over the $S^2$. It degenerates at some  radius~$r_+$ when $F$ vanishes, in which case:
\begin{equation}
\label{eq:mass-parameter}
m=\frac{r_+^2+n^2}{2r_+}+\frac{1}{2\ell^2}\left(r_+^3-6n^2 r_+ - 3\frac{n^4}{r_+}\right)\ .
\end{equation}
To ensure that there is no conical singularity, we have 
$F^\prime(r=r_+)=1/2n$. The parameters $m$  and $n$, and the range of $r$   have certain allowed ranges that depend upon whether we are Taub--NUT or Taub--Bolt, and the conditions have been worked out in the literature\cite{Chamblin:1998pz}:

\bigskip

\noindent
$\bullet$ For the Taub--NUT case $r_+=r_n=n$. The spacetime there is locally $\mathbb{R}^4$, a ``nut''\cite{Gibbons:1979xm}. The mass parameter $m=m_n$, and the equation~(\ref{eq:mass-parameter}) simplifies to:
\begin{equation}
\label{eq:taub-nut-mass}
m_n=n-\frac{4n^3}{\ell^2}\ ,
\end{equation}
and we have that $n\leq r \leq +\infty$. 

\bigskip

\noindent
$\bullet$ For the Taub--Bolt case $r_+=r_b>n$, and it satisfies a quadratic equation:
\begin{equation}
\label{eq:arbyquadratic}
6nr_b^2-\ell^2 r_b -6n^3+2n\ell^2=0\ ,
\end{equation}
with an upper ($r_{b+}$) and lower ($r_{b-}$) branch solution:
\begin{equation}
\quad  r_{b\pm}=\frac{\ell^2}{12 n }\left(1\pm\sqrt{1-48\frac{n^2}{\ell^2}+144\frac{n^4}{\ell^4}}\right)\ ,
\end{equation}
with $n$ restricted to be no greater than $n_{\rm max}$ (where the branches join) in order  to have real $r_b$ greater than $n$:
\bigskip
\begin{equation}
n_{\rm max}=\left(\frac16-\frac{\sqrt{3}}{12}\right)^{\frac12}\ell\ .
\end{equation}
The form of the function $r_b(n)$ is  plotted in figure~\ref{fig:arbyplot}. 

\bigskip
\noindent The spacetime in the neighbourhood of $r=r_b$ is locally $\mathbb{R}^2\times S^2$, a ``bolt''\cite{Gibbons:1979xm}. In that case the mass parameter $m=m_b$, where we use the mass formula~(\ref{eq:mass-parameter}) with $r_+=r_b$, 
and we have that $r_b\leq r \leq +\infty$. The upper and lower branches are analogous to the large and small black hole branches  possessed by AdS--Schwarzschild black holes\cite{Hawking:1982dh}. (The computations are structurally different, but with similar results. For black holes the horizon radius $r_+$ is given and then consistency yields a $\tau$ period $\beta(r_+)$. Here, the period $8\pi n$ is given, and then consistency\footnote{With the temperature $T=1/8\pi n$ for AdS--Taub--NUT/Bolt, one might wonder about the case $n=0$ which reduces the metric~(\ref{eq:metric}) back to AdS--Schwarzschild. The latter has a temperature that depends explicitly on the mass. There is no problem with the $n\to0$ limit, where for AdS--Taub--NUT/Bolt, $T\to\infty$. One should also follow the  mass formulae~(\ref{eq:mass-parameter}) and~(\ref{eq:taub-nut-mass}), which show that $m$ vanishes or (for the large $r_{b+}$ branches) diverges as $n\to0$. This perfectly matches  the  $T\to\infty$ behaviour of the small and large black hole AdS--Schwarzschild solutions that the $n=0$ metric yields    for those mass values.} produces a function $r_b(n)$.) The two branches'  characteristics become sharply distinct for small $n$ (equivalently large $\ell$), where
\begin{equation}
r_{b-}\simeq 2n+\frac{18n^3}{\ell^2}+O(n^5)\ , \qquad r_{b+}\simeq \frac{\ell^2}{6n}-2n-\frac{18n^3}{\ell^2}+O(n^5)\ ,
\end{equation}
showing that the lower branch, constituted of bolts whose radii are ``small'' compared to the length scale $\ell$ set by the cosmological constant, connect smoothly to the asymptotically locally flat ($\Lambda=0$) case.

\section{Thermodynamic Variables}
\label{sec:thermo-variables}
The temperature of these spacetimes is set by the inverse of $\tau$'s period, $T=1/8\pi n$. The action of each  was computed in refs.\cite{Mann:1999pc,Emparan:1999pm} and may be written in the  very simple form given 
in ref.\cite{Emparan:1999pm}:
\begin{equation}
I=\frac{4\pi n }{G\ell^2}\left(\ell^2m+3n^2r_+-r_+^3\right)\ .
\end{equation}
 For Taub--NUT we put $m=m_n$ and $r_+=n$  while for  the Taub--Bolt we use $m=m_b$ and $r_+=r_b$. The difference between the actions of the two spacetimes  (which was actually computed before the expression above was known, by using a subtraction method\cite{Chamblin:1998pz,Hawking:1998ct}), can be written as:
\begin{equation}
\label{eq:action_diff}
I_b-I_n= -\frac{2\pi n}{G\ell^2}\left(\frac{(r_b-n)^2(r_b^2-6nr_b-n^2)}{r_b-2n}\right)\ .
\end{equation}
The point here is that it is natural, at a given $n$, to compare the action of Taub--NUT to that of Taub--Bolt since they have the same asymptotics at large $r$.  
 The previous work on the phase structure derived by examining this action difference was done at fixed cosmological constant. In the extended thermodynamics $\Lambda=-3/\ell^2$ defines a pressure $p=3/(8\pi G\ell^2)$ and so we have instead a line of phase transition points. We will discuss this in a later section.

 The entropy can be derived from the action using $S=(\beta\partial_\beta-1)I$, and after substituting to make the pressure more explicit we have the form\cite{Johnson:2014xza}:
\begin{equation}
\label{eq:entropy}
 S = \frac{4\pi n}{G}\left[\frac{r_+^2+n^2}{2r_+}+\frac{4\pi Gp}{3}\left(3r_+^3 -12 n^2 r_+-\frac{3n^4}{r_+}\right)\right]\ .  \nonumber
\end{equation} 
Note that there is in principle $\ell$ dependence and hence pressure dependence in the $r_+$ function as well. It is only in the case of Taub--NUT, where $r_+=r_n=n$, that things simplify to give a very  simple $p$ dependence, or in the case of small bolts, where $r_+\to 2n$ and we return in the limit to zero cosmological constant. We will discuss these cases below.

Finally,  in the extended thermodynamics it was proposed in ref.\cite{Johnson:2014xza} that the mass for the Taub--NUT and Taub--Bolt spacetimes define the enthalpy of the system, and so we write: $H=m/G$:
\begin{equation}
 H(S,p)= \frac{1}{G} \frac{r_+^2+n^2}{2r_+}+\frac{4\pi}{3}p\left(r_+^3-6n^2 r_+ - 3\frac{n^4}{r_+}\right)\ ,
\end{equation} 
and this results in an interesting expression for the conjugate to pressure, the thermodynamic volume\cite{Johnson:2014xza}:
\begin{equation}
\label{eq:thermodynamic-volume}
V = \frac{4\pi}{3}\left(r_+^3-3n^2 r_+\right)\ .
\end{equation}
Note that this thermodynamic volume is (in general) not to be associated directly with any geometric volume in the spacetime, as discussed in detail in ref.\cite{Johnson:2014xza}.  It can also be negative, which was also given  an interpretation in ref.\cite{Johnson:2014xza}: the environment did positive work in forming the spacetime, as opposed to having work done on it, as is more familiar with an enthalpy. 

The volume for Taub--NUT is $\ell$--independent and so is also true in asymptotically locally flat space  (the limit of $p=0=\Lambda$). Interestingly, the result for the Taub--Bolt spacetime in asymptotically flat space (using at large $\ell$ the  small branch family: $r_{b-}\to2n$ in the limit) is equal in magnitude to that of Taub--NUT, with opposite sign:
\begin{equation}
\label{eq:special-volume-limit}
\left.V_b\right|_{p=0} = \frac{8\pi n^3}{3} = -V_n\ .
\end{equation}
The NUT and Bolt spacetimes seem  rather complementary in their roles. In this limit, it is worth noting the other (finite) quantities as well\cite{Hunter:1998qe}:
 \begin{eqnarray}
 \label{eq:zero_pressure}
 &&\biggl.H_n\biggr|_{p=0} \equiv \biggl.\frac{m_n}{G}\biggr|_{p=0} = \frac{n}{G}\ ; \qquad \biggl.I_n\biggr|_{p=0} =  \frac{4\pi n^2}{G}\ ; \qquad \biggl.S_n\biggr|_{p=0} =  \frac{4\pi n^2}{G}\ . \nonumber \\
 &&\biggl.H_b\biggr|_{p=0} \equiv \biggl.\frac{m_b}{G}\biggr|_{p=0} = \frac{5}{4}\frac{n}{G}\ ; \qquad \biggl.I_b\biggr|_{p=0} =  \frac{5\pi n^2}{G}\ ; \qquad \biggl.S_b\biggr|_{p=0} =  \frac{5\pi n^2}{G}\ . 
 \end{eqnarray}

\section{Phase Structure}
\label{sec:phases}

\begin{wrapfigure}{R}{0.4\textwidth}
{\centering
\includegraphics[width=2.6in]{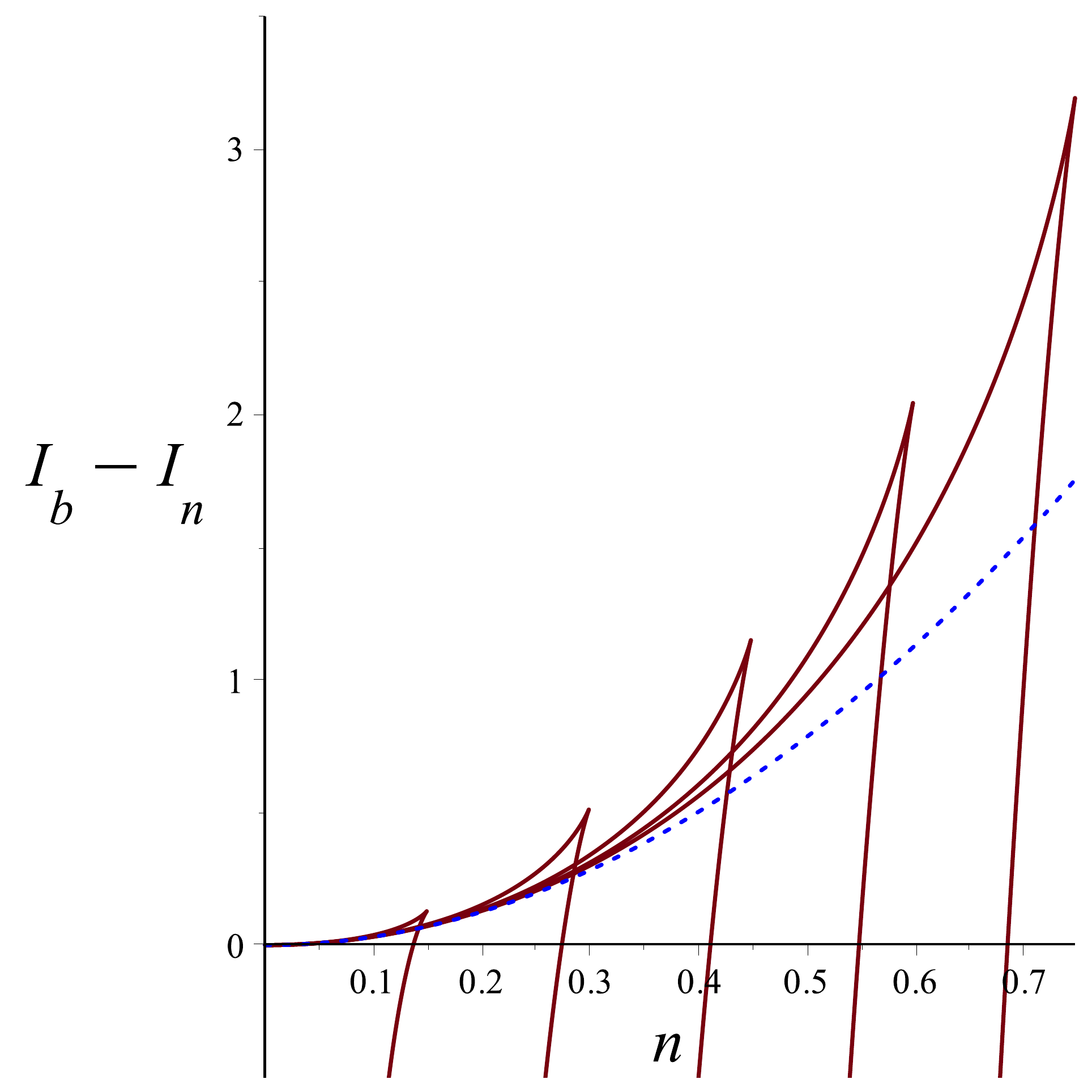} 
   \caption{\footnotesize    The action (or Gibbs free energy) difference for  pressures $p=3/(8\pi\ell^2 G)$, for $\ell=1,2,3,4,5$, with $G=1$. The curves move more to the right as the pressure decreases. The blue dotted line  is the case of zero pressure,  which the solid curves approach in the limit.}  \label{fig:action_difference}
}
\end{wrapfigure}
The phase structure of the Taub--NUT/Taub--Bolt system, as studied in ref.\cite{Chamblin:1998pz,Hawking:1998ct}, was only the structure at a particular value of the cosmological constant, and hence at fixed pressure. 
The action difference between each spacetime was studied as a function of temperature. At a certain temperature, the Taub--Bolt spacetime had lower action, and so there was a first order transition where the difference $I_b-I_n$ in equation~(\ref{eq:action_diff}) becomes negative. Since there are two branches of bolts, the curve of $I_b-I_n$ as a function of $n$ (which sets the inverse temperature) is made of two parts that meet (at $n_{\rm max}$) to form a cusp shape. The transition is for the large bolts (the $r_{b+}$ branch).   

In the extended thermodynamics where pressure is a variable, we can explore the phase transitions found in the earlier work further, seeing that they form a line of points in the ($p,T$) plane.  Each value of the pressure  $p=3/(8\pi\ell^2 G)$ gives a cusp curve where the cusp starts at $n_{\rm max}$ and proceeds on two branches to lower $n$. In the limit of infinite~$\ell$ (zero pressure), it is only the part of the cusp that corresponds to the small bolts (the $r_{b-}$ branch) that has finite action, and for them $I_b-I_n=\pi n^2/G$. See figure~\ref{fig:action_difference} and equation~(\ref{eq:zero_pressure}). (Parenthetically, notice that the curves can all be superimposed by a scaling. This actually follows from  equation~(\ref{eq:action_diff}) for the action difference. If $n\to\alpha n$ and $\ell\to\alpha\ell$, then $(I_b-I_n)\to\alpha^2(I_b-I_n)$. This all results from  the fact that an accompanying scaling of the coordinates ($\tau\to\alpha\tau$ and $r\to\alpha r$) rescales\footnote{We thank an anonymous referee for highlighting this scaling.} the overall metric of equation~(\ref{eq:metric}): $ds^2\to\alpha^2 ds^2$.)

The action difference~(\ref{eq:action_diff}) changes sign at the positive zero of the quadratic  in the numerator, which is at $r_b=(3+\sqrt{10})n$. Putting this into the quadratic equation~(\ref{eq:arbyquadratic}) satisfied by $r_b$ gives an equation for the value of $n$ at which the transition happens, which we denote as $n^*$. After some algebra,
\begin{equation}
n^*=\frac16 (\sqrt5-\sqrt2)\ell\ ,
\end{equation}
and  further algebra can be used to turn this into  an expression for the transition temperature~$T^*$, as a function of pressure:
\begin{equation}
\label{eq:transition-temp}
T^*(p)=\frac{\sqrt{3G}}{\sqrt{2\pi}(\sqrt{5}-\sqrt{2})}p^\frac12 \ .
\end{equation}
\begin{wrapfigure}{R}{0.43\textwidth}
{\centering
\includegraphics[width=2.6in]{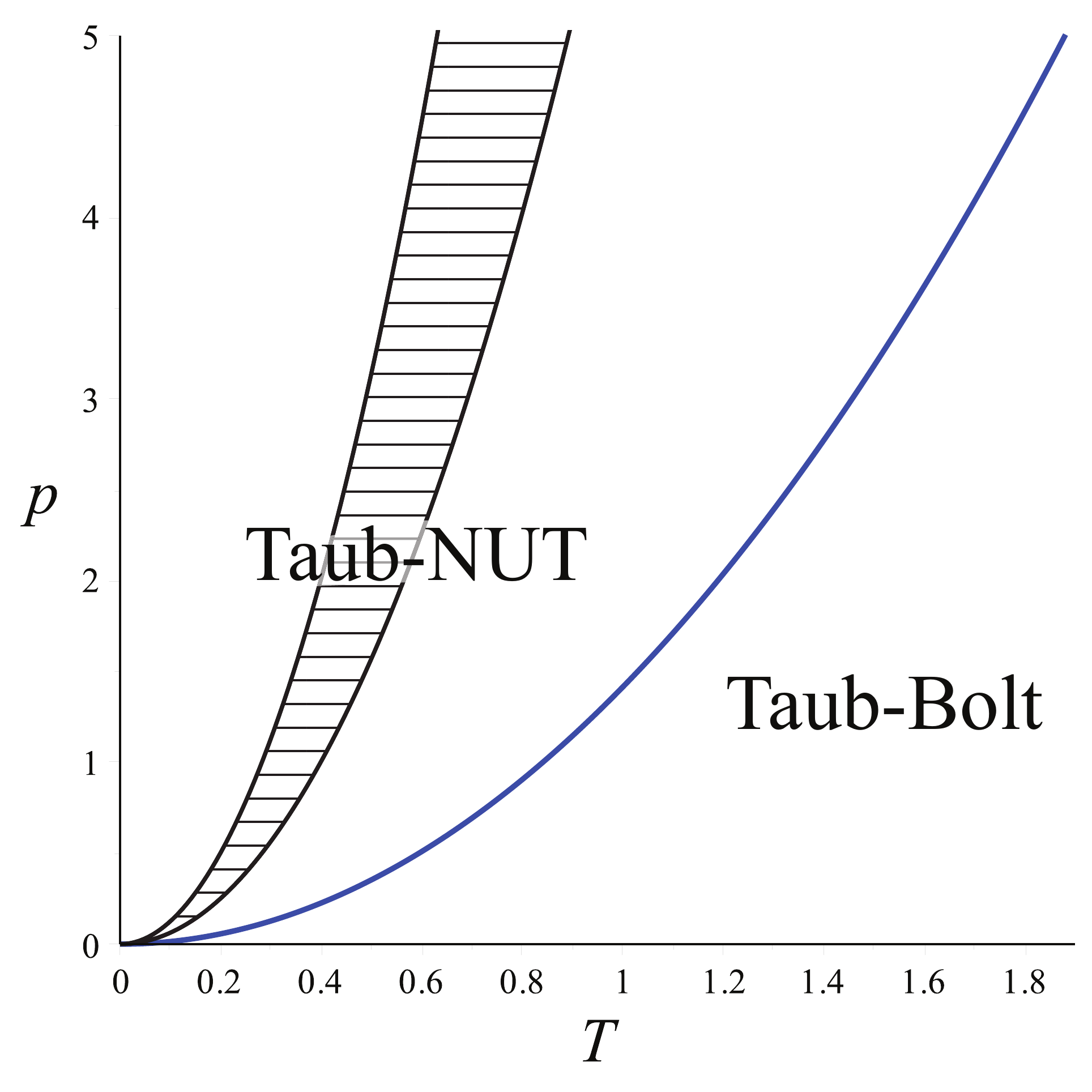} 
   \caption{\footnotesize   The rightmost solid curve represents the line of transition between the Taub--NUT and Taub--Bolt phases. The shaded region between the two  curves on the left, referred to as the ``positive wedge'' in the text, is where entropy $S>0$ and specific heat $C_p>0$. For points in the region between the positive wedge and the transition, the system is dynamically driven by fluctuations toward the transition line.}  \label{fig:transitiontempplot}
}
\end{wrapfigure}
So far this is an analogue of a solid--liquid coexistence curve\footnote{This $T\sim p^{1/2}$ form of the curve, and other phase boundaries in the $(p,T)$ plane that we will uncover,  follows from the scaling properties  discussed at the end of the previous paragraph, since $T\to T/\alpha$ and $p\to p/\alpha^2$.} in the phase diagram, although further study in this section will show that only one phase can exist near the line, due to a thermal instability. Then section~6 will study a deformation that will allow for regimes where it is genuinely a coexistence line. 

The $p=0$ point on the curve is deceptive. In fact, strictly at zero pressure, there is no bolt phase at all, as the large bolts have gone to infinite size, and their action is infinite. But we see now in this larger context that this is a very special point that neighbours on a richer story. The transition temperature is
lower for lower pressure, and so as soon as there is some non--zero cosmological constant (pressure) there will be a nearby transition to a bolt solution.
 
This is analogous to the Hawking--Page black hole transition, where  at finite cosmological constant there is a transition from pure AdS to an AdS--Schwarzschild black hole, but that transition disappears at zero cosmological constant: Flat space Schwarzschild black holes (the small branch survivors of the two branches available at finite $\Lambda$) have higher action than flat space at any temperature. In fact, the varying pressure coexistence curve can be computed in that case too (see ref.\cite{Altamirano:2014tva} for a recent review), with the same functional dependence, but with coefficient $\sqrt{8/3\pi}$ instead of the one displayed in equation~(\ref{eq:transition-temp}) for our case.

The diagram has two additional curves plotted on it as well. The shaded region between them has positive values for both the entropy and the specific heat at constant pressure, $C_p=T\left.\partial S/\partial T\right|_p$. We'll call this the ``positive wedge'' in the $(p,T)$  phase diagram, for the purposes of discussion. Along a  fixed pressure slice, for low enough temperatures, the entropy is negative, while at high enough 
 temperatures $C_p<0$. Between these two temperatures, both are positive. The temperatures can be read off from  equation~(\ref{eq:entropy}) specialised to the Taub--NUT entropy (by putting $r_+=n$) to give:
 \begin{equation}
 \label{eq:entropy_nut}
 S_n=\frac{4\pi n^2}{G}(1-16\pi G p n^2)\ , 
 \end{equation}
 from which the specific heat can be computed as\footnote{The latter result follows since the Taub--NUT thermodynamic volume  $|V_n|=8\pi n^3/3$ depends strictly on the temperature only through $T=1/8\pi n$ and hence fixing the  volume fixes $S$. This is analogous to the vanishing of $C_V$ for static black holes, although their origins are quite different. There, the entropy and volume are not independent quantities, both depending upon a single parameter, the radius $r_h$ of the black hole.}
  \begin{equation}
 \label{eq:specific_heat_nut}
 C_p=\frac{8\pi n^2}{G}(32\pi G p n^2-1)\ ,\quad C_V = 0\ . 
 \end{equation}

\begin{wrapfigure}{R}{0.45\textwidth}
{\centering
\includegraphics[width=3.0in]{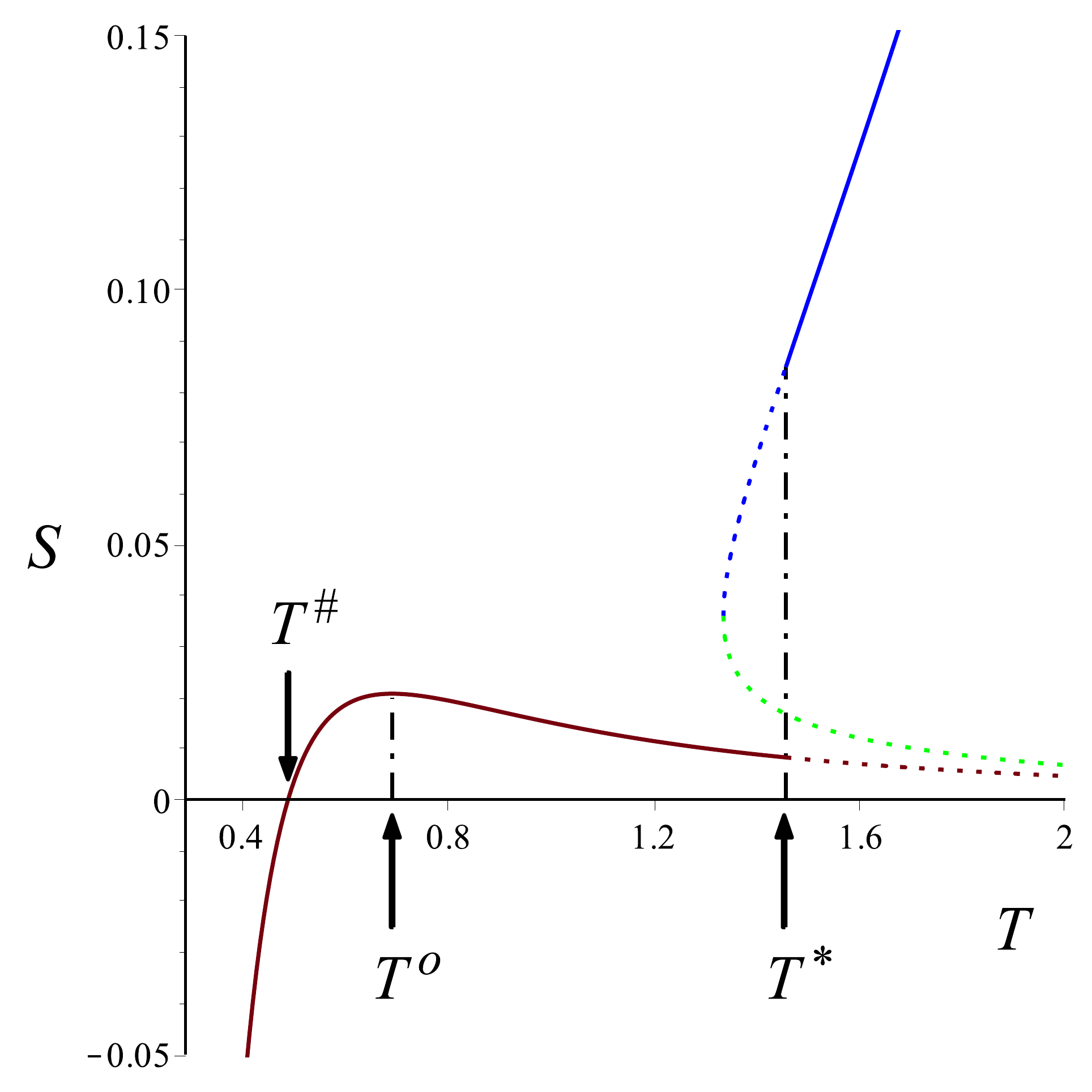} 
   \caption{\footnotesize   The entropy curves  as a function of temperature, at fixed pressure. Here $p=3$.  As $T$ increases, there is a transition from Taub--NUT to Taub--Bolt at~$T^*$. Dotted lines show solutions with higher action.  $T^\#$ and $T^o$ are discussed in the text.}  \label{fig:entropyplot}
}
\end{wrapfigure}

\bigskip
\noindent
From the above expressions we can compute the profile of the  curves that border the positive wedge. At a given value of the pressure $p$, we have for the lower temperature:
\begin{equation}
 \label{eq:entropy_region_lower}
 T^\#\equiv\frac{1}{8\pi n^\# }=\left(\frac{G}{4\pi}\right)^\frac12 p^\frac12\ ,
 \end{equation}
 and for the upper temperature:
 \begin{equation}
 \label{eq:entropy_region_upper}
T^o\equiv\frac{1}{8\pi n^o}=\left(\frac{G}{2\pi}\right)^\frac12p^\frac12\ ,
 \end{equation}
and so we have $T^\# \leq T\leq  T^o$ for the slice of the positive wedge located at pressure $p$. After the transition temperature $T^*$, the entropy of the favoured  bolt solutions is both positive and has $C_p>0$. Figure~\ref{fig:entropyplot} shows the entropy as a function of $T$ for a fixed pressure slice, and the three temperatures $T^\#,T^o,T^*$.

The negative entropy of Taub--NUT for the region $T< T^\# $ and the negative specific heat for~$T$ in the region between $T^o$ and the transition   at $T^*$  are identified as   puzzles in refs.\cite{Mann:1999pc,Emparan:1999pm}. There is a very natural narrative in the context here, {\it i.e.,} extended thermodynamics. For a start, having a negative entropy for the spacetime can quite naturally arise as a result of some net heat outflow during the same formation process that accompanies the interpretation of  the mass as enthalpy of formation, as discussed in ref.\cite{Johnson:2014xza}. We won't have any more to say about that here, except to note that we can, if we choose to, avoid the negative entropy region altogether by appropriate paths in the $(p,T)$ plane. For example, one can start at $p=0$ and turn on finite temperature. The entropy is positive and given in equations~(\ref{eq:zero_pressure}). The same set of equations shows that the action of Taub--NUT is lower than Taub--Bolt  and so is thermodynamically favoured. Then one can directly enter the positive wedge by turning on small enough pressure~$p$.  We can move around inside the positive wedge and do  standard equilibrium thermodynamics. Notice that the special case when Taub--NUT is actually pure AdS$_4$ with an $S^3$ slicing, $n=\ell/2$, {\it i.e.} $T=\sqrt{(G/6\pi)}p^{1/2}$, lies outside this wedge.

Now we wish to consider how to traverse the region in figure~(\ref{fig:transitiontempplot}) between the positive wedge in the Taub--NUT region and the transition at $T^*$ to Taub--Bolt. Are there any obstructions to doing so? The fact that $C_p$ goes negative at the edge of the wedge signals what presumably happens. The system develops an instability to fluctuations above the edge temperature $T^o=\sqrt{(G/2\pi)} p^{1/2}$, and it  is dynamically predisposed to  drive itself  to higher temperature, where the specific heat becomes even more negative, improving the susceptibility to instability. This happens until the system arrives at $T^*$ and transitions to Taub--Bolt, where it is again stable.

We can go further and track the changes in the system through this process in the extended thermodynamics. It is easy to do at fixed pressure, {\it i.e.,} moving along a state curve of the form given in figure~\ref{fig:entropyplot}. This isobaric process has a volume change, and so (in the spirit of ref.\cite{Johnson:2014yja}) we should take this seriously as  a physical process during which work is done either on or by the system, with an accompanying heat flow. This  changes  the internal energy  of the system, which was computed in ref.\cite{Johnson:2014xza} to be:
\begin{equation}
\label{eq:taub_nut_energy}
U_n = \frac{n}{G}\left(1-8\pi G p n^2\right)\ .
\end{equation} All of this is of course accounted for using the First Law of thermodynamics, naturally written in terms of the enthalpy $H_n=U_n+pV_n$ as $dH_n=T dS_n +V_ndp$. Our isobar gives us:
\begin{eqnarray}
\Delta H_n= \int_{n^o}^{n^*}\frac{1}{8\pi n}\frac{\partial S_n}{\partial n} dn &=& \frac{1}{G} (n^*-n^o) - \frac{32\pi p}{3}[(n^*)^3-(n^o)^3]\nonumber\\
&=&\frac{1}{G} (n^*-n^o) - {8\pi p}[(n^*)^3-(n^o)^3]  - p\frac{8\pi }{3}[(n^*)^3-(n^o)^3]\nonumber\\
&=&\Delta U_n+p\Delta V_n\ .
\end{eqnarray}

In summary then,  it is natural to think of the pressure as a parameter that allows us to explore a larger phase space of parameters that deform the Taub--NUT/Bolt system  away from flat space (the origin). A critical line develops separating Taub--NUT and  Taub--Bolt phases, and two regions  develop in the Taub--NUT area, one has $S<0$ and the other has $C_p<0$, separated by the ``positive wedge''. We can avoid the $S<0$  region by appropriate choice of path in the $(p,T)$ plane. Fluctuations dynamically drive points in the $C_p<0$ region toward the transition line to become Taub--Bolt solutions. Presumably this means that one cannot quasi--statically follow a path from the Taub--Bolt region back toward Taub--NUT solutions\footnote{So if constructing holographic heat engines using these spacetimes as a ``working substance'', as done for black holes in ref.\cite{Johnson:2014yja}, the cycles cannot include paths that traverse the $C_p<0$ region.}. Furthermore this means that the transition line is strictly not a phase coexistence line  like those familiar in solid/liquid transitions. The $C_p<0$ region means that the nuts and bolts, despite having equal action (Gibbs free energy) there, do not coexist indefinitely. Taub--Bolt dominates after a long enough time.

However, the relative  locations of the positive wedge region and the transition line are adjustable using a natural deformation parameter. We explore that in the next two sections, finding that there are some values of the parameter where the transition line is more like a standard coexistence line.

\section{Dyonic Taub--NUT--AdS and  Taub--Bolt--AdS}
The Taub--NUT/Bolt spacetime has many interesting properties. One of them  is that while it possesses ordinary mass, it also has nut charge,  which can in fact be interpreted as a magnetic counterpart to the ordinary ``electric'' mass. In this sense it is a dyon, for the mass sector, and this helps contribute to Taub--NUT's role as a self--dual gravitational instanton\cite{Hawking:1976jb}, where the Euclidean section sets the mass $m$ equal to the nut charge $n$. 

One might wonder about the fate of Taub--NUT if one embeds it non--trivially into electromagnetism, extending the Einstein--Hilbert action to incorporate Maxwell:
\begin{equation}
I=-\frac{1}{16\pi G}\int \! d^4x \sqrt{-g} \left(R-2\Lambda -F^2\right)\ ,
\label{eq:actionM}
\end{equation}
with $F$ (not to be confused with the metric function in equation~(\ref{eq:metric})) the standard gauge field  two--form with components $F_{\mu\nu}=\partial_\mu A_\nu-\partial_\nu A_\mu$.
It was shown in ref.\cite{PhysRev.133.B845} that there is a charged geometry, and in fact  both an electric and magnetic field are turned on. We have a dyon in the electromagnetic sense as well\footnote{In fact, it is natural to study dilatonic versions as well, for example for non--trivial embedding into string theory, as first done using exact heterotic string conformal field theory techniques in ref.\cite{Johnson:1994jw}, and using T-- and S--duality transformations in ref.\cite{Johnson:1994ek}. The result is that there an axion charge turns on as well, and so the system becomes triply dyonic, with the dilaton and axion being the electric/magnetic pairs. We will not study the addition of such fields in this paper.}.  A more general solution can be written down that has non--zero cosmological constant, and it has the same form for the metric given in equation~(\ref{eq:metric}), but with a more general form for the metric  function $F(r)$ given by\cite{Plebanski1975196,Plebanski:1976gy,AlonsoAlberca:2000cs}:

\begin{equation}
\label{eq:metric-functionM}
F(r)\equiv\frac{(r^2+n^2+4n^2v^2-q^2)-2mr+\ell^{-2}(r^4-6n^2r^2-3n^4)}{r^2-n^2}\ ,
\end{equation}
and the gauge sector is:
\begin{equation}
A\equiv A_{\mu}dx^{\mu}=h(r)(d\tau-2n\cos\theta d\phi)\ ,
\end{equation}
where:
\begin{equation}
h(r)= \frac{qr}{r^2-n^2}+v\frac{r^2+n^2}{r^2-n^2}\ .
\end{equation}

Conditions of smoothness of the Euclidean section will result in the parameter $q$ being related to the parameter $v$ (analogous to $m$ being related to $n$ in the familiar neutral case), giving us a   deformation from the uncharged system. Setting them to zero recovers the solutions we discussed above.

The analogous conditions for the smoothness of the Euclidean section were discussed in refs.\cite{Mann:2005mb,Awad:2005ff}, and some  thermodynamic quantities discussed there and in ref.\cite{Dehghani:2006dk}, including (for the latter case) a computation of the action  for the charged Taub--NUT and Bolt cases. We will not review all this here, since the ideas are very similar to those we presented in an earlier section, with the addition of the condition that the gauge potential $A$ must be regular at the position of the nut or bolt at $r=r_+$ where $F(r_+)=0$. We simply list below some of the results that we need, in order to discuss the phase structure (which has not been explored in the literature). The mass can be written as:
\begin{equation}
\label{eq:mass-parameter-deformed}
m=\frac{r_+^2+n^2+4n^2v^2-q^2}{2r_+}+\frac{1}{2\ell^2}\left(r_+^3-6n^2 r_+ - 3\frac{n^4}{r_+}\right)\ ,
\end{equation}
with 
\begin{equation}
\label{eq:charge}
q=-v\frac{r_+^2+n^2}{r_+}\ .
\end{equation}
The electric charge  and potential at infinity given by
\begin{equation}
\label{eq:charge_and_potential}
Q=q\ , \qquad \Phi = -\frac{q r_+}{r_+^2+n^2} = v\ .
\end{equation}
\begin{wrapfigure}{L}{0.45\textwidth}
{\centering
\includegraphics[width=2.6in]{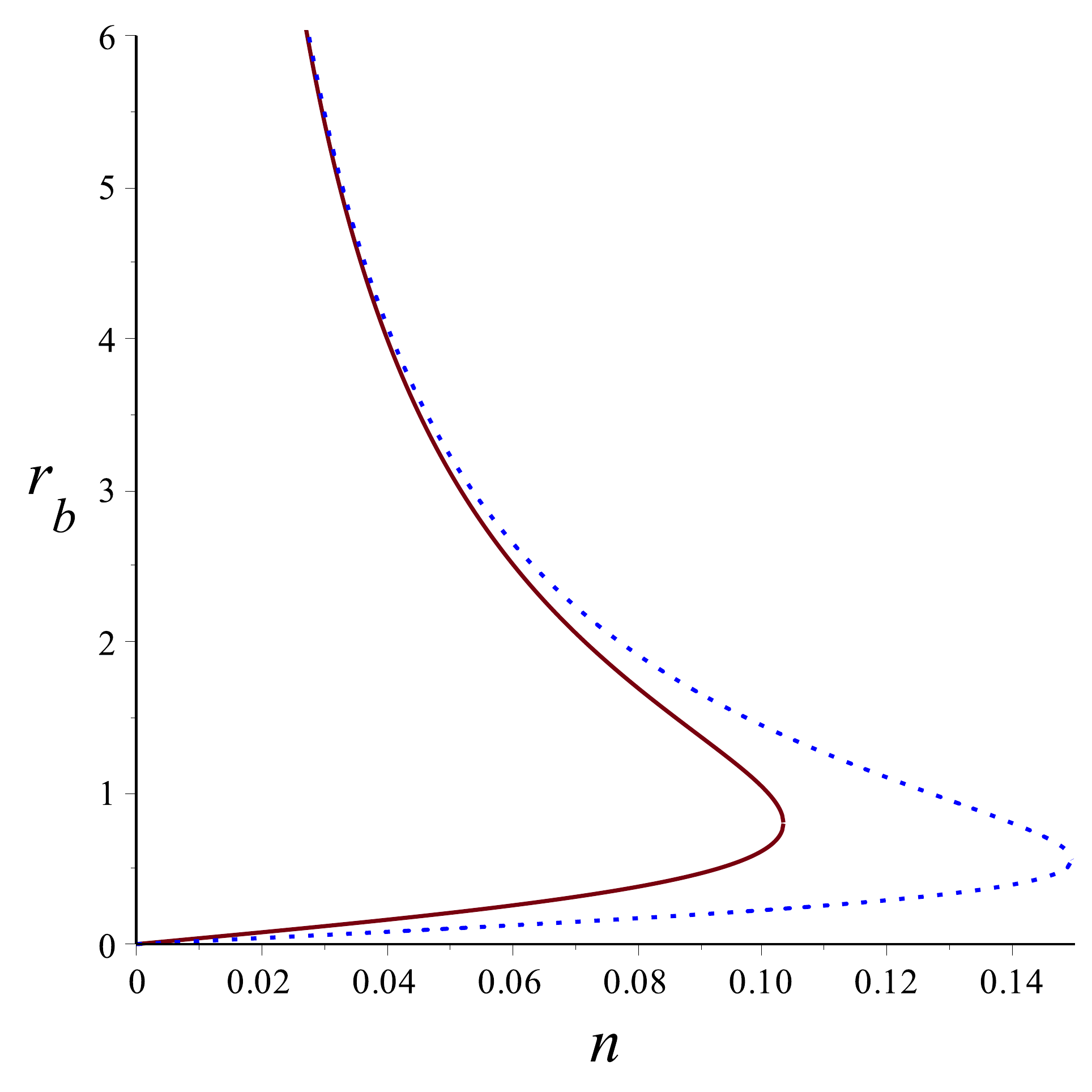} 
   \caption{\footnotesize   The available dyonic bolt radii, the union of an upper and a lower branch that connect at $n_{\rm max}^{(v)}\simeq 0.103\ell$. Here we chose $\ell=1$ and $v=1$. The dotted line is the neutral case for comparison. See text for discussion.}  \label{fig:arbyplot_dyonic}
}
\end{wrapfigure}

\noindent
The solution is a dyonic nut if $r_+=r_n=n$, and a dyonic bolt if $r_+=r_b$ satisfies the fourth order equation:
\begin{equation}
\label{eq:arbyquartic}
6nr_b^4-\ell^2 r_b^3 +2n(\ell^2-3n^2+\ell^2v^2)r_b^2 - 2\ell^2v^2n^3=0\ .
\end{equation}
This is a more complicated condition than our quadratic in equation~(\ref{eq:arbyquadratic}). One  might wonder if there is potential for certain kinds of new structure here, in analogy with what happens for charged black holes. In that case, the more complicated equation for the horizon radius in the presence of charge produced a new physical branch of solutions at small horizon radius and high  temperature that gave a new phase structure in the fixed charge ensemble\cite{Chamblin:1999tk,Chamblin:1999hg}. The line of critical points in the phase diagram ended in a new second order critical point.  One might wonder if that can happen here. Indeed, there are two branches of new solutions that appear, but they do not add (in the Euclidean section) any new physical structures: Taking the small $r_b/\ell$ limit of  equation~(\ref{eq:arbyquartic}) gives the cubic $r_b^3-2n(1+v^2)r_b^2+2v^2n^3=0$. At $v=0$ the two new branches are the repeated (unphysical) root $r_b=0$, joining our familiar large $\ell$ solution $r_b=2n$. Turning on  $v$ then shows that the two branches grow out as  a pair of roots from the origin of the  $(r_b,n)$ plane, one positive, the other negative. It can be seen that both have  $r_b<n$, and so are not physical.

Instead we have a system of physical solutions that is similar in structure to what we saw before (see figure~\ref{fig:arbyplot_dyonic} for an example; {\it c.f.} figure~\ref{fig:arbyplot}). There are two branches of bolts, a ``small'' and a ``large'', and they join at a deformed value of a maximum~$n$ we can call $n_{\rm max}^{(v)}$, which is less than $n_{\rm max}$. Correspondingly, we might expect that the transition from the dyonic Taub--NUT to the dyonic Taub--Bolt is at lower~$n$, or higher temperature, and we will confirm this below.

Since there are no new branches of solution for $r_b$ in the presence of charge, it seems that it qualitatively does not matter too much whether we use the fixed charge or fixed potential ensemble to study the phase structure although of course there are differences in the precise values of the parameters at transition. We can sketch out what happens, picking the fixed potential ensemble\footnote{This is the easier ensemble to study in comparing the NUT and Bolt cases since fixing $v$ is easier than fixing $q$ when comparing the two solutions,  because (looking at equation~(\ref{eq:charge})) $q$ depends upon $r_b$, which is  a complicated function of $v$.}.

The action of the system was computed in ref.\cite{Dehghani:2006dk} using an extension of the techniques outlined in refs.\cite{Balasubramanian:1999re,Emparan:1999pm}, with the result (in the fixed potential ensemble):
\begin{equation}
I=-\frac{2\pi}{G}\frac{[r_+^4-\ell^2r_+^2+n^2(3n^2-\ell^2)]r_+^2-(r_+^4+4n^2r_+^2-n^4)\ell^2v^2}{(3r_+^2-3n^2+\ell^2)r_+^2+(r_+^2-n^2)\ell^2v^2}\ .
\end{equation}
The entropy was computed as:
\begin{equation}
S=\frac{2\pi}{G} \frac{[3r_+^4+(\ell^2-12n^2)r_+^2+n^2(\ell^2-3n^2)]r_+^2}{(3r_+^2-3n^2+\ell^2)r_+^2+(r_+^2-n^2)\ell^2v^2}+
\frac{2\pi}{G}\frac{(r_+^4+4n^2r_+^2-n^4)\ell^2v^2}{(3r_+^2-3n^2+\ell^2)r_+^2+(r_+^2-n^2)\ell^2v^2}\ .
\label{eq:dyon-entropy}
\end{equation}

Let us pause to see how this all looks in the case of the dyonic nut, where the expressions are much simpler. We can put $r_+=n$ to get:
\begin{equation}
m_n=n-\frac{4n^3}{\ell^2}\ ,\quad Q= -\frac{2nv}{G}\ ,\quad \Phi = v\ , \end{equation}
and
\begin{equation}
\label{eq:deformed-entropy}
I_n=\frac{4\pi n^2}{G}\left(1-\frac{2n^2}{\ell^2}+2v^2\right)\ , \quad
S_n=\frac{4\pi n^2}{G}\left(1-\frac{6n^2}{\ell^2}+2v^2\right)\ .
\end{equation}
Interestingly, the mass is the same as in the neutral case. This means that the enthalpy is unchanged in the extended thermodynamics, which leads to another new result: A quick computation shows that the thermodynamic volume $V_n$ for the dyonic Taub--NUT stays the same as it was for the neutral case. Put differently, the shift in the entropy is entirely accounted for by the new $-Q\Phi/2$ term in the Smarr\cite{Smarr:1972kt}  relation:
\begin{equation}
\frac{H}{2}-TS+pV-\frac{Q\Phi}{2}= 0\ .
\end{equation}
Just as in ref.\cite{Johnson:2014xza} for the neutral case, we can use the same Smarr relation to deduce the thermodynamic volume for the dyonic bolt case. The resulting expression does depend (explicitly and implicitly through $r_b$'s dependence on $v$) on the deformation parameter $v$ and it is long and not useful for us here, so we  will not write it.

\section{Deformed Phase Structure}
\label{sec:deformed_phases}
From our Taub--NUT--dyon's entropy, we see that the specific heat at constant pressure is now:
\begin{equation}
\label{eq:deformed-specific}
C_p^{(v)} = -\frac{8\pi n^2}{G}\left(1-\frac{12n^2}{\ell^2}+2v^2\right)\ .
\end{equation}
\begin{wrapfigure}{R}{0.39\textwidth}
{\centering
\includegraphics[width=2.5in]{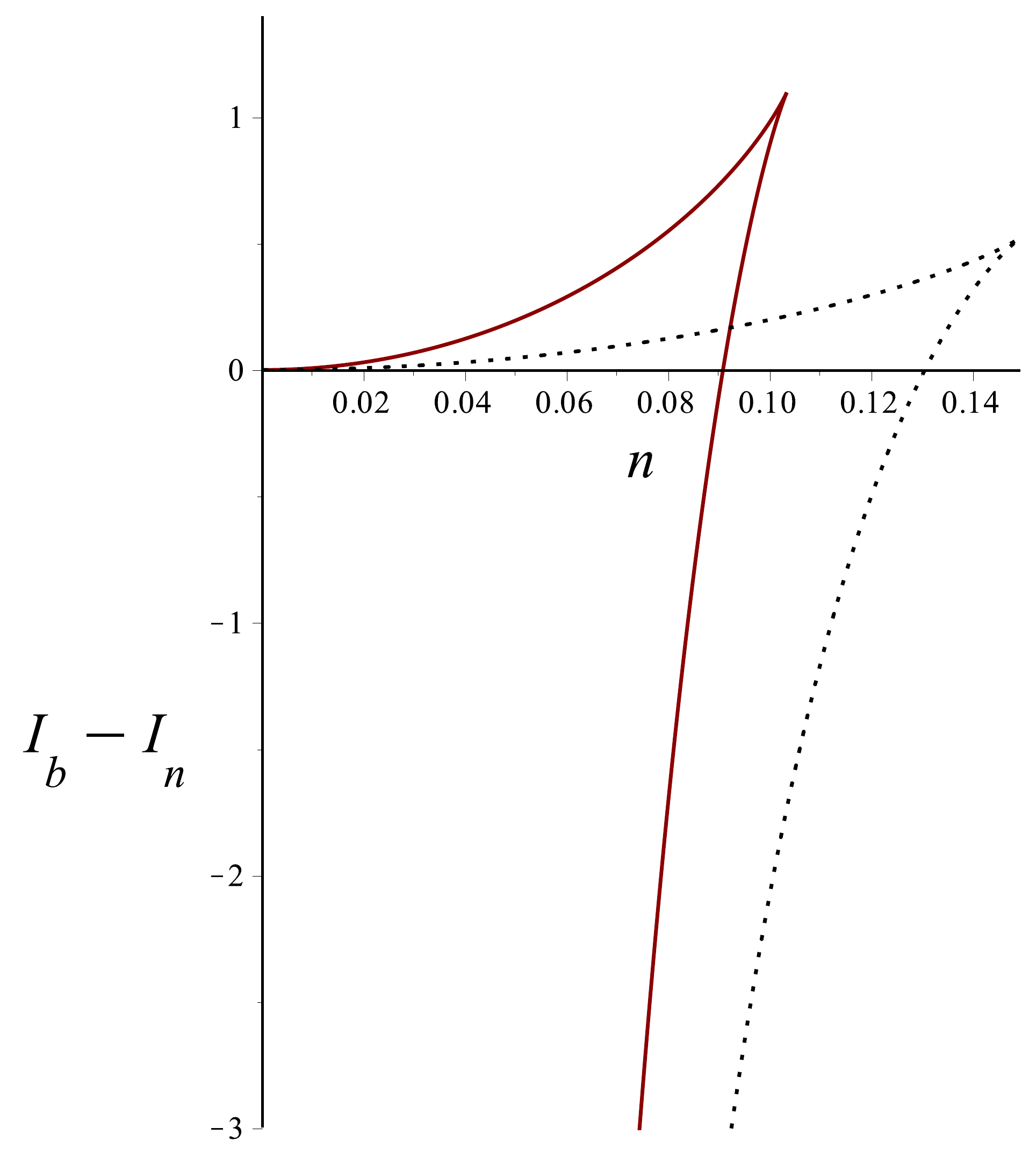} 
   \caption{\footnotesize   The solid curve is the difference between the action for dyonic Taub--Bolt and dyonic Taub--NUT in the fixed potential ensemble. The dotted curve is the neutral case, for comparison. The transition temperature has increased. Here we used $\Phi=v=1$, and $\ell=1$.}  \label{fig:dyonic_action_difference}
}
\end{wrapfigure}

So  the deformation parameter $v$ has shifted the location of the place in the $(p,T)$ plane  where the entropy is negative, and where the specific heat goes negative. Our previously obtained temperatures for the borders of the positive wedge get rescaled by a factor $(1+2v^2)^{-1/2}$. The temperature where the entropy starts going positive is lower, as well the temperature where the specific heat $C_p$  becomes negative. The dyonic charge evidently hastens the destabilisation of Taub--NUT. In the limit of infinite $v$,  entropy becomes positive for all temperatures, but then  $C_p<0$ everywhere.

The phase structure for the dyonic system can now be explored, although analytic expressions do not seem to be possible for all regions.  A numerical exploration reveals some  pleasant results, however. For example, figure~\ref{fig:dyonic_action_difference} shows a plot of the action difference between dyonic Taub--Bolt and dyonic Taub--NUT as a solid curve, against the neutral case  for comparison (a dashed curve). They have the same value of the pressure ($\ell=1$ and we've chosen units such that $G=1$) and $v=1$ for the dyon. As anticipated,  the transition temperature has been raised.

\begin{figure}[ht]
{\centering
\subfigure[Real dyonic  deformation]{\includegraphics[width=2.6in]{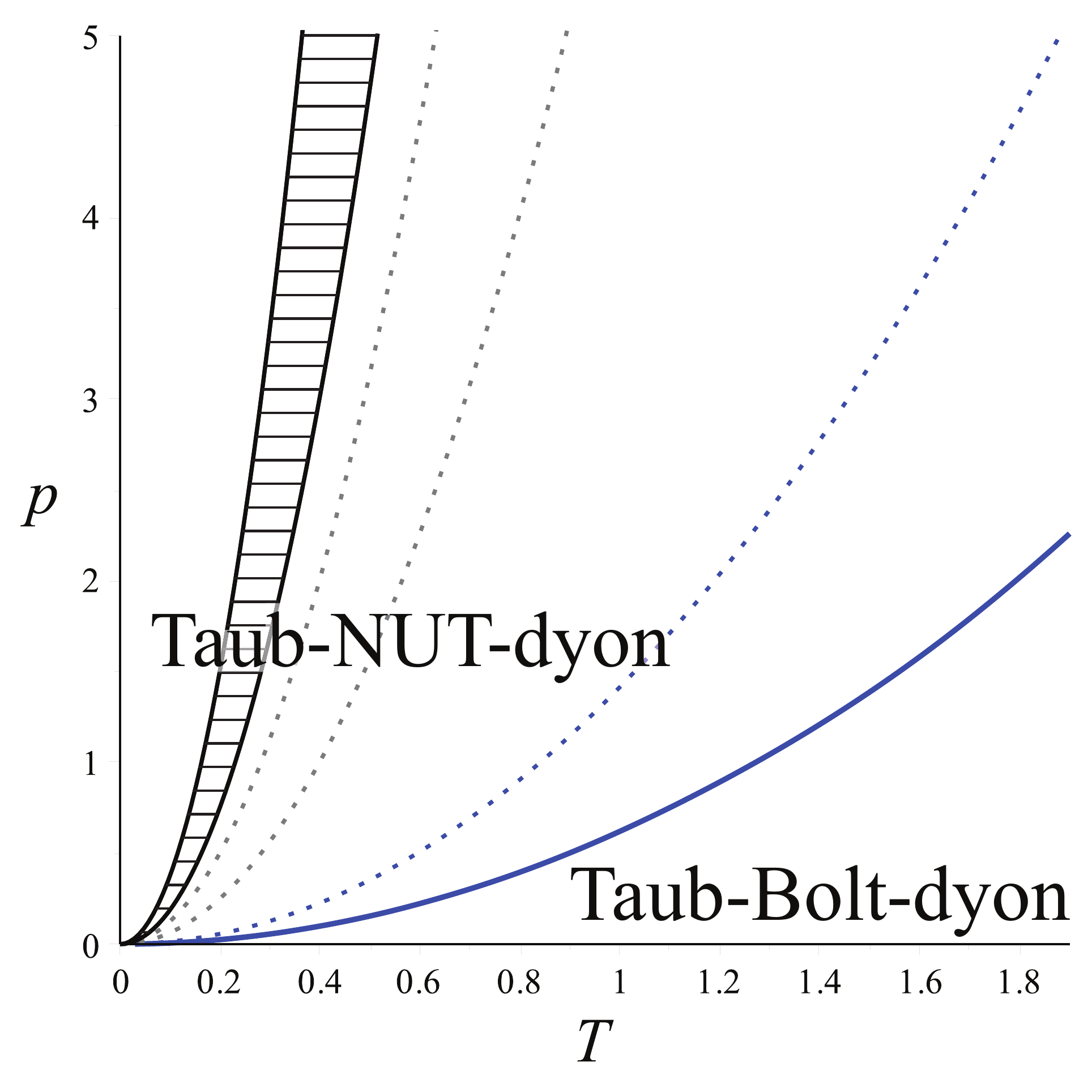}  \label{fig:transitiontempplotdyona}}\hspace{1.5cm}
\subfigure[Imaginary dyonic  deformation]{\includegraphics[width=2.6in]{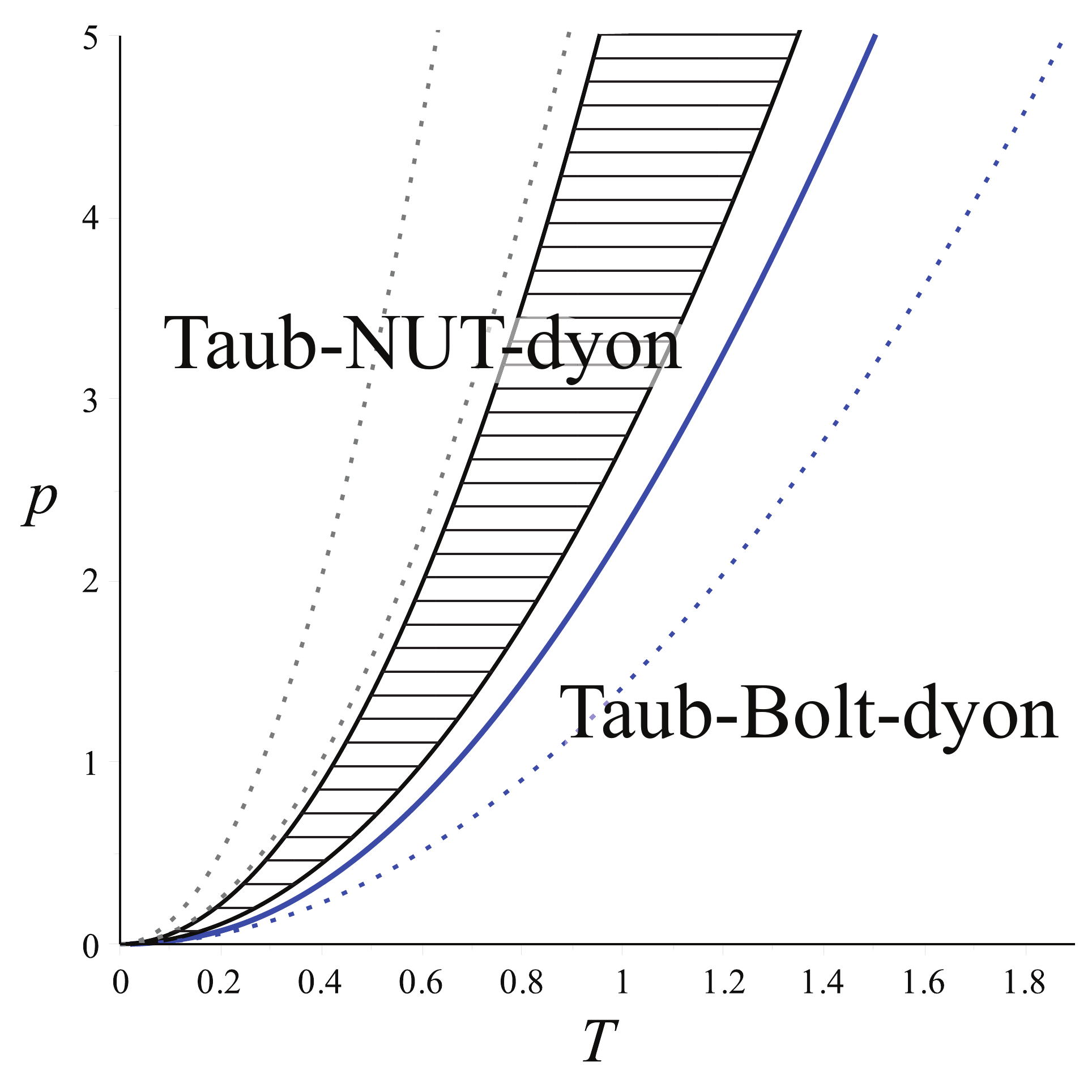}  \label{fig:transitiontempplotdyonb}}
   \caption{\footnotesize  The phase  diagram for the Taub--NUT--dyon and Taub--Bolt--dyon phases, for (a) real dyon deformation parameter $v^2=1$, and (b) complex dyon parameter $v^2=-9/32$. The  ``positive wedge'' (shaded region)  has shifted left and the (solid) transition line  has tilted right. See text for explanation.  The dotted lines are for comparison to the neutral case in figure~\ref{fig:transitiontempplot}.}} 
\end{figure}

A series of such examples can be generated numerically for varying pressure, extracting the transition temperature for each case. This  allows for the generation of  a new phase diagram for the dyonic deformation as a companion to the neutral one we generated in figure~\ref{fig:transitiontempplot}. The result is shown in figure~\ref{fig:transitiontempplotdyona}. There seems to be no simple analytic expression for the shape of the first order  transition line for non--zero $v$, although it fits the $T^2$ dependence rather well. 

The deformation widens the size of the unstable $C_p<0$ region. The fate of the phase diagram in the large $v$ limit is therefore interesting, since in that  case the $C_p<0$ region increasingly envelops the phase diagram, at the expense of the negative entropy region, which shrinks away, along with the positive wedge, against the $p$--axis. 

Although its full meaning is unclear, it is worth noting that the case where we continue~$v$ to being purely imaginary is also interesting\footnote{Here, we are perhaps in good company: The Lee--Yang edge singularity, an important model in  the study of phase transitions, arises from continuing the Ising model in an external magnetic field to regimes where the magnetic field is imaginary\cite{Yang:1952be,Lee:1952ig}. Imaginary $v$ here corresponds to imaginary magnetic and electric fields for our dyons.}.  From equations~(\ref{eq:deformed-entropy}) and~(\ref{eq:deformed-specific}),  we see that  the positive wedge is now pushed more toward the right,  toward the transition line. A numerical exploration shows that the temperature at which there are available bolt solutions is decreased  ({\it i.e.} $n_{\rm max}^{(v)}$ is now greater than $n_{\rm max}$). Correspondingly, the first order  phase transition line tilts more toward the left, and so approaches the positive wedge for increasing imaginary $v$. The $C_p<0$ region for Taub--NUT gets squeezed. See figure~\ref{fig:transitiontempplotdyonb}. 

\begin{figure}[h]
{\centering
\subfigure[Imaginary dyonic  deformation entropies]{\includegraphics[width=2.6in]{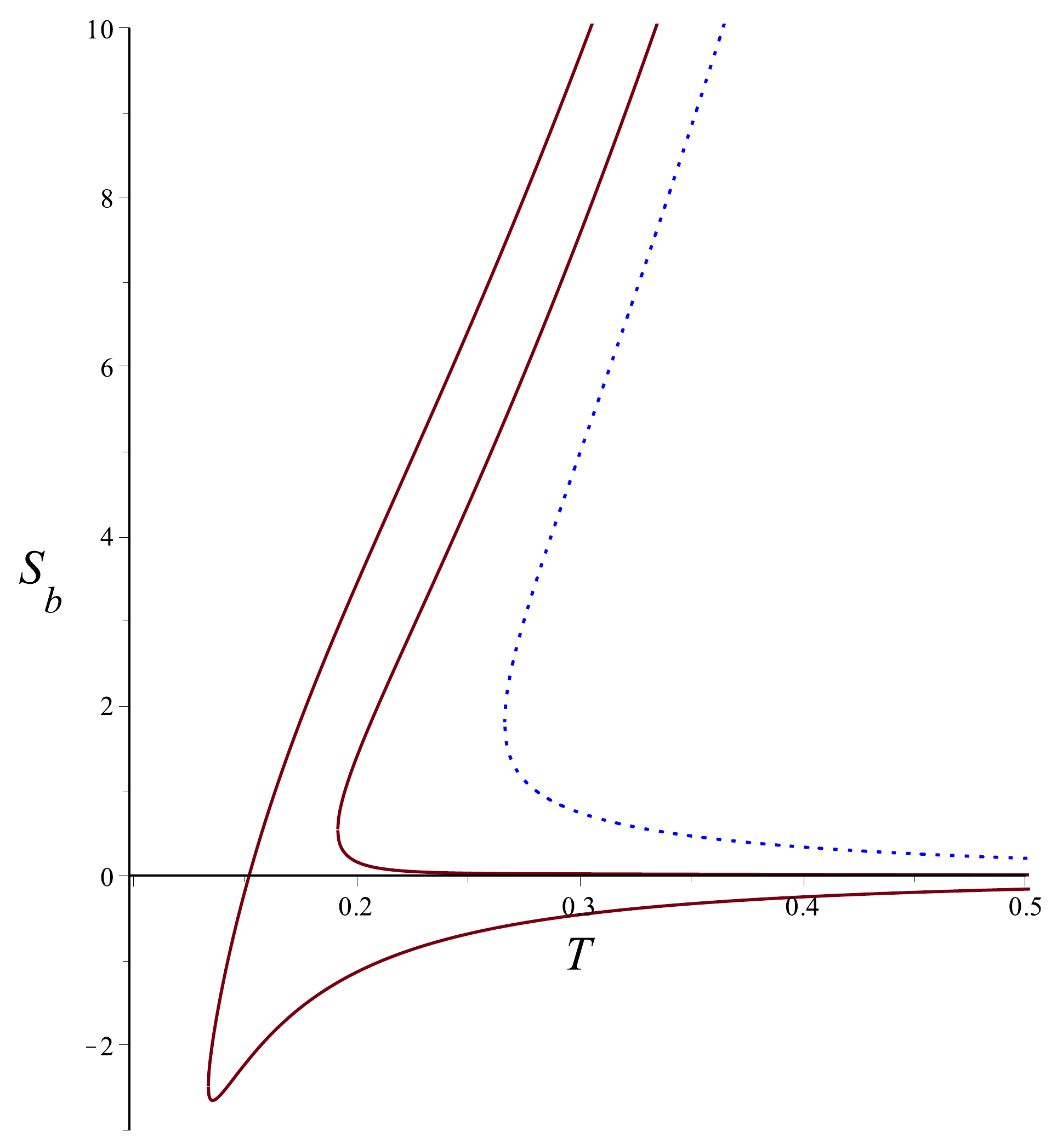}  \label{fig:complex-dyon-snapshotsa}}\hspace{1.5cm}
\subfigure[Imaginary dyonic  deformation actions]{\includegraphics[width=2.6in]{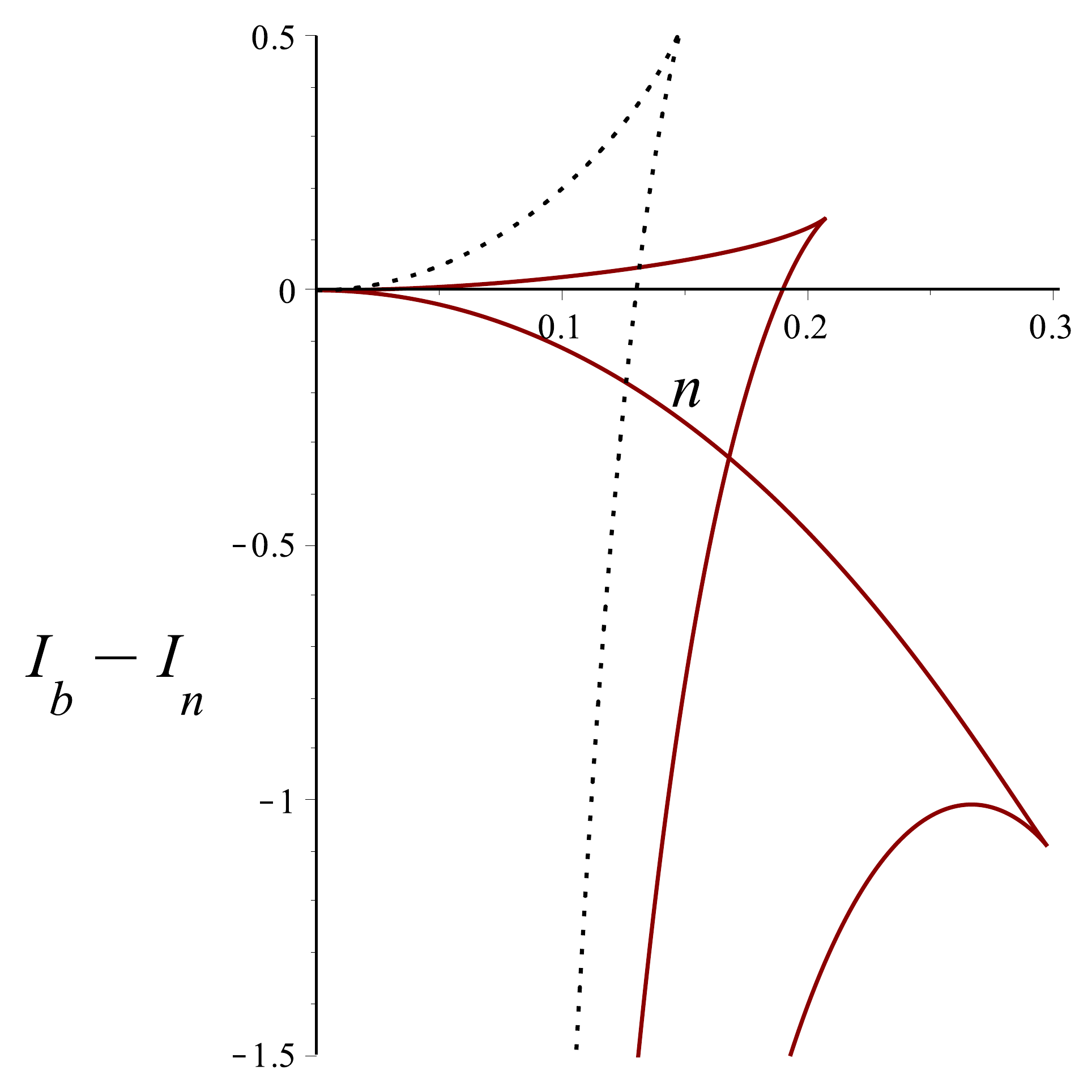}  \label{fig:complex-dyon-snapshotsb}}
   \caption{\footnotesize  Some examples of curves showing the dyonic Taub--Bolt entropy ({\it vs.} $T$) and action difference  ({\it vs.} $n$). The dotted lines are for comparison to the neutral case. We have $G=\ell=1$ in all plots. In plots (a), the curves, reading  from left to right, have $v^2=-1,-1/2,0$, which are the values of the curves reading right to left in plots (b).}} 
\end{figure}

Ultimately there are yet  new  regimes that appear (for increasingly large imaginary $v$). One occurs when the $C_p<0$ region  no longer has lowest action.  In other words, the transition line to Taub--Bolt is inside the positive wedge: $T^*<T^o$. In this case the transition line is genuinely a coexistence line analogous to that found in solid/liquid systems. Another regime is when (at $v^2=-1/2$) the $C_p<0$ Taub--NUT region vanishes entirely. The whole positive wedge disappears, leaving a phase diagram where there is only Taub--NUT with negative entropy before the transition line to Taub--Bolt. Interestingly,   yet another new phase can  appear at even higher imaginary $v$, as illustrated by the samples in figures~\ref{fig:complex-dyon-snapshotsa} and~\ref{fig:complex-dyon-snapshotsb}. 

The large Taub--Bolt solutions themselves  begin to develop regions of negative entropy (as is evident from the form of equation~(\ref{eq:dyon-entropy})). The sample plots of the action difference show that for  large enough~$|v|$ the large Taub--Bolt solutions become thermodynamically favoured as soon as they become available, as their action relative to the Taub--NUT solutions are negative for all $n<n_{\rm max}^{(v)}$. So these negative entropy Taub--Bolts appear on the phase diagram for a window of temperatures, until at higher temperatures the positive entropy Taub--Bolt solutions take over again.   Perhaps  there is even more  interesting and instructive physics to explore further in this regime.



\section*{Acknowledgements}
 CVJ would like to thank the  US Department of Energy for support under grant DE-SC0011687,  and Amelia for her support and patience.


\providecommand{\href}[2]{#2}\begingroup\raggedright\endgroup

\end{document}